\documentclass[12pt,preprint]{aastex}
\newcommand{\Msun}{~M_\odot}
\newcommand{\Rsun}{~R_\odot}

\newcommand{\kms}{\rm ~km~s^{-1}}

\newcommand{\ergs}{\rm ~erg~s^{-1}}
\newcommand{\ml}{~\Msun ~\rm yr^{-1}}

\begin{document}

\title{TYPE IIb SUPERNOVAE WITH COMPACT AND EXTENDED PROGENITORS}
\author{
Roger A. Chevalier\altaffilmark{1} and
Alicia M. Soderberg\altaffilmark{2,3}
}
\altaffiltext{1}{Department of Astronomy, University of Virginia, P.O. Box 400325, 
Charlottesville, VA 22904-4325; rac5x@virginia.edu}
\altaffiltext{2}{Harvard-Smithsonian Center for Astrophysics,
60 Garden St., MS-51, Cambridge, MA 02138}
\altaffiltext{3}{Hubble Fellow
}

\begin{abstract}
The classic example of a Type IIb supernova is SN 1993J, which had a cool extended progenitor surrounded by a dense wind.  There is evidence for another category of Type IIb supernova which has a more compact progenitor with a lower density, probably fast, wind.  Distinguishing features of the compact category are: weak optical emission from the shock heated envelope at early times; nonexistent or very weak H emission in the late nebular phase; rapidly evolving radio emission; rapid expansion of the radio shell; and expected nonthermal as opposed to thermal X-ray emission.  Type IIb supernovae that have one or more of these features include
SNe  1996cb, 2001ig, 2003bg, 2008ax, and 2008bo.  All of these with sufficient radio data (the last four) show
evidence for presupernova wind variability.  We estimate a progenitor envelope radius $\sim1\times 10^{11}$ cm for SN 2008ax, a value consistent with a compact Wolf-Rayet progenitor.  Supernovae in the SN 1993J extended category include SN 2001gd and probably the Cas A supernova.  We suggest that the compact Type IIb events be designated Type cIIb and the extended ones Type eIIb.
The H envelope mass dividing these categories is $\sim0.1\Msun$.
\end{abstract}


\section{INTRODUCTION}

A Type IIb supernova has a spectrum with high velocity H lines near maximum, but in the late
nebular phases more closely resembles a Type Ib supernova, with weak or absent H lines.
The first supernova placed in this category was SN 1987K \citep{filippenko88}.
However, it was the nearby SN 1993J in M81 that has come to best define the properties
of a SN IIb.
In this case, the progenitor star was identified as a K supergiant \citep{aldering94},
consistent with the interpretation of the early light curve which required
a stellar radius of $\sim 4\times 10^{13}$ cm \citep{woosley94}.
Such a star is expected to have a slow, dense wind, and there is radio and X-ray
evidence for a wind with mass loss rate $\dot M\approx 4\times 10^{-5}\ml$ for an
assumed wind velocity of $v_w=10\kms$ \citep{fransson96,fransson98}.
Radio VLBI observations provide strong support for the wind interaction picture
\citep{marcaide97,bartel02}.
The interaction with the dense wind also produces some H$\alpha$ emission in the
late nebular phase \citep{matheson00}.

A consistent scenario for SN 1993J is thus the explosion of an extended star that
interacts with the dense wind from the progenitor star.
However, there is also evidence for Type IIb supernovae arising from compact stars,
and that is the topic addressed here.
Some of that evidence has already been presented for SN 2001ig \citep{ryder04},
SN 2003bg \citep{soderberg06}, and SN 2008ax \citep{pastorello08}.
In \S~2, we present the differences between SNe IIb that have extended or
compact progenitors.
The results are discussed 
in \S~3.

\section{TYPE IIb SUPERNOVA CATEGORIES}

\subsection{The Progenitor Star and Early Light Curve Evolution}

The most direct way to observe the progenitor star is in presupernova observations.
This has been possible for SN 1993J, as discussed above, and has recently been
possible for SN 2008ax.
\cite{crockett08} find that the progenitor of the Type IIb SN 2008ax is either a single, He-rich,
Wolf-Rayet star or a stripped star in an interacting binary in a low mass cluster.

The light curve evolution in the first days after the explosion, which is the time
of cooling envelope emission after shock breakout, provides a good
indicator of the initial stellar extent.
The cooling envelope emission is more luminous for a more extended star because of the 
larger emitting area and adiabatic expansion losses are less important for an
extended star.
The early peak luminosity and first 10 days of evolution provided evidence for
a large radius, $\sim 4\times 10^{13}$ cm for the progenitor of SN 1993J \citep{woosley94}.

In the case of SN 2008ax, a comparable early luminous phase was absent in the light
curve \citep[][Fig. 11]{pastorello08}.
Very early observations with {\it Swift} did show evidence for an initial peak
in the $uvw1$, $u$ and $b$ bands \citep{roming09}.
The luminosity level of the emission is comparable to, although slightly fainter than,
similar emission observed in the Type Ib/c SN 1999ex \citep{stritz02}.
We used the cooling envelope theory of \cite{chevalier08} to model the $uvw2,uvw1,u,b$ emission
on day 1.7 \citep{roming09} and the $g,r,i,z$ emission on day 1.9 \citep{pastorello08}.
The magnitude results of \cite{roming09} were converted to fluxes using
the flux conversion factors, pivot wavelengths, and bandpass
widths for the {UVOT} as reported in \cite{poole08}.
For a distance of 9.6 Mpc, a blackbody fit to the fluxes gives
an effective temperature $T\approx 6000$ K and a photospheric radius
$r_{ph}=4.3\times 10^{14}$ cm on day 1.8.
Taking an ejecta mass of $M_{ej}=3.8\Msun$ and an energy $E=1.5\times 10^{51}$ ergs
from the model of \cite{tsvetkov09}, equation (2) of \cite{chevalier08} yields
$r_{ph}=3.8\times 10^{14}$ cm, in good agreement with the observed value.
Equation (5) of \cite{chevalier08} and the observed $T$ yield a progenitor
radius of $R=1\times 10^{11}$ cm. 
However, the temperature is sufficiently low that recombination is becoming important,
so that the model of \cite{chevalier08} is inaccurate.
The progenitor radius estimate should still be valid to within a factor $2-3$,
showing that the progenitor was compact, with a radius consistent with
a Wolf-Rayet star.
The radius determined here is that of the stellar envelope; the photospheric 
radius may be a factor $2-3$ larger due to the stellar wind \citep[see Fig. 1 of][]{li07}.
Although the explosion parameters for SN 2008ax 
are similar to SN 1993J \citep{pastorello08,tsvetkov09}, 
the progenitor is very different, as indicated by the progenitor observations and the early light curve.

For a compact star, the cooling envelope emission lasts for only a few days, so
that observations must be carried out very soon after the explosion for this phase
to be detected.
Observations that rule out an extended progenitor may be the most that can be obtained.
For the Type IIb SN 1996cb, there are  light curve observations showing that an early luminous phase
like that of SN 1993J was not present \citep{qiu99}.

\subsection{Circumstellar Interaction}

Circumstellar interaction depends on the progenitor radius in that more compact
stars have faster, and thus more rarefied, winds and shock acceleration at the time
of breakout extends to higher velocities in compact stars.
Radio emission is an excellent indicator of circumstellar interaction and Fig. 1
shows peak radio luminosity vs. time of peak for SN IIb and SN Ib/c.
SSA (synchrotron self-absorption) is expected to be the dominant absorption mechanism for the high velocities
present in SN Ib/c and can be used to estimate the radio shell
velocity at the time of the peak radio luminosity \citep{chevalier98}.
The SN IIb designation has been given to SN 1993J, SN 2001gd \citep{matheson01},
SN 2001ig \citep{phillips01}, SN 2003bg \citep{hamuy09}, and SN 2008ax \citep{pastorello08}.
Fig. 1 shows that there is a class of SN IIb that have radio properties similar
to those of SN Ib/c, i.e. they have rapid radio evolution for a given luminosity,
indicating velocities of $30,000-50,000\kms$.
These velocities are expected for the explosion of a Wolf-Rayet star because of
strong shock acceleration in the outer parts of the star and the low wind density resulting
from a high wind velocity \citep{chevalier06}.
They are designated here as SN cIIb because of the expectation
that they have relatively compact progenitors.
There is direct light curve and progenitor evidence for this in the case of SN 2008ax, as discussed above.
SN 1993J is placed in the SN eIIb category because of its extended progenitor, as is
SN 2001gd, because of its radio properties.
In addition to the radio light curve information on SN 2001gd \citep{stockdale07},
there are also VLBI observations on days  286 and  582 from explosion, which combine
to give a mean velocity
$\sim 9100\kms$ on day 434 with 20\% uncertainty \citep{perez05}.
This velocity is consistent with that found from the radio light curves (Fig. 1)
and is somewhat less than that observed for the interaction shell in SN 1993J at
a comparable time, $\sim 14,000\kms$ \citep{bartel02,marcaide97}.

VLBI observations of SN 2008ax \citep{marti09} indicate a velocity $\sim 52,000\kms$ (in a plausible 
model) at an age of 33 days, supporting the SSA model.
The highest H velocity in optical lines was only about half this, as noted
by \cite{marti09}.
This is not a problem for the Type cIIb case because the high
velocity shocked H is nonradiative, as discussed below.
The unshocked gas is of such low density that it is not seen.
There do not seem to be sufficient early radio observations of SN 2008ax \citep{roming09}
to test for SSA from the light curves, but in the case of SN 2003bg, there is support
for the SSA model \citep{soderberg06}.

The SSA model of \cite{chevalier06} can be used to obtain estimates of the
mass loss properties of the compact supernovae.
We characterize the mass loss density density by $A_*$, where
$\dot M/v_w=A_* \times 10^{-5}\ml/1000\kms$ and find
$A_*\epsilon_{B-1}\alpha^{8/19}=3$ for SN 2001ig, $13$ for SN 2003bg,
and $0.65$ for SN 2008ax; here,
$\epsilon_{B-1}$ is the fraction of the postshock energy density in magnetic field
in units of 0.1
and $\alpha$ is the ratio of relativistic electron energy density to magnetic energy density.
The results for SN 2001ig and SN 2003bg are the same as in \cite{chevalier06}
and that for SN 2008ax uses the  observations of \cite{roming09}.
Assuming that the efficiency of production of the radio emission is comparable for
the different objects, the variation in the radio luminosities of the supernovae
translates to a variation in the progenitor mass loss densities, by a factor of 20.

Other Type IIb supernovae have been detected in the radio, but are not included in
Fig. 1 because the available data do not allow the maximum luminosity to be
accurately determined.   
However, there is some information.
SN 2008bo showed optical properties similar to SN 2008ax, including deep
H$\alpha$ absorption \citep{navas08}.
There were early radio detections of the supernova and a later (age $\sim20$ days)
rebrightening \citep{stockdale08}.
At a distance of 21 Mpc, the radio luminosity was $\sim 1.5$ times lower than
that of SN 2008ax and the emission became optically thin somewhat earlier,
so this object extends the apparent wind density to slightly lower values.
There are also 2 radio observations of the Type IIb SN 1996cb that indicate a $\tau=1$
frequency of $\sim 11$ GHz on Dec. 21, 1996 \citep{vandyk96}.
The observations indicate that it also lies on the $L_p - t_p$
plot close to SN 2008ax.

X-ray emission is another signature of circumstellar interaction.
\cite{chevalier06} find that the X-ray emission from SNe Ib/c is probably due
to a nonthermal mechanism: inverse Compton emission near optical maximum light
and synchrotron emission at later times.
The general expectation is that Type cIIb events should have nonthermal
X-ray emission, while Type eIIb events similar to SN 1993J should have
dominant thermal X-ray emission, as in SN 1993J \citep{fransson96}.
There are not yet sufficiently well observed Type cIIb events to test this
from the spectra themselves.
However, \cite{roming09} found a decrease in the X-ray luminosity of SN 2008ax
from $6.0\pm1.9\times 10^{38}\ergs$ in the first month to 
$1.4\pm0.9\times 10^{38}\ergs$ in the second month.
This is the magnitude of decline that would be expected in going from an inverse
Compton component near optical maximum to a synchrotron component
a month later \citep[Fig. 1 of][]{chevalier06}.

Finally, another signature of a dense circumstellar medium is optical emission
from gas heated and ionized by radiation from
reverse shock heated gas \citep{chevalier94}.
The optical emission can be either from freely expanding ejecta or from
a dense shell formed as a result of radiative cooling.
SN 1993J did show evidence for persistent H$\alpha$ emission to an age
of 2500 days \citep{matheson00}; the box-like line profile and other characteristics
were consistent with circumstellar interaction.
The early H$\alpha$ emission from SN 1993J showed a decline consistent with a
radioactive power source; the late steady phase started at an age of $300-350$ days
\citep{houck96}.
The estimated wind density in SN 1993J corresponds to $A_*=400$ \citep{fransson96}, 
which is a factor $>30$ larger than the wind densities deduced here for the SNe cIIb.
Cooling is primarily by bremsstrahlung and the reverse shock is nonradiative
over the times of observations for SN cIIb.
The low density and large shock radius imply that H$\alpha$ from circumstellar interaction
should be undetectable in the late nebular phase.
However, \cite{stritz09} suggest that circumstellar interaction powers H$\alpha$
emission in the nebular phase from the Type Ib SN 2007Y, even though upper limits on the radio
emission yield $A_*\la 0.1$.
The explanation is probably that the observations are on days 230 and 270, when
radioactivity is still a plausible power source.
Observations of H$\alpha$ at times $\ga 1$ yr are needed to distinguish between
the steady emission that can occur in SNe  eIIb  and the continued drop
expected in SNe cIIb.

\subsection{Structure in Wind}

Both SN 2001ig \citep{ryder04} and SN 2003bg \citep{soderberg06} showed structure
in their radio light curves that was interpreted as density structure in the
wind region.
\cite{ryder04} suggested that the structure was due to a binary companion, as
has been observed in the pinwheel dust structure around some Wolf-Rayet
stars.
\cite{ryder06} in fact found evidence for a possible companion star at the site of the supernova.
However, the star is of type late-B to late-F; such a star would have too weak
a wind to have much effect on the strong wind deduced for the progenitor of SN 2001ig.
\cite{soderberg06} argued for stellar variability rather than a binary companion,
based on the similarity between the radio evolution of SN 2001ig and SN 2003bg.

The additional cases of Type cIIb supernovae also show evidence for wind density variations.
The radio light curves of SN 2008ax show a peak in the optically thin phase at $t\approx 60$ days
\citep{roming09}.
In the case of SN 2008bo, there is a possible re-brightening at $t\approx 20$ days
\citep{stockdale08}.
There is thus evidence for density variations in every proposed Type cIIb supernova for
which there are sufficient radio observations to show the effect.
On the other hand, the Type Ib event SN 2008D \citep{soderberg08} and the Type Ic
events SN 2003L \citep{soderberg05} and SN 2009bb \citep{soderberg09} show smooth light
curves.
For SN 2003bg, \cite{soderberg06} estimated that the time between the higher density
outflows was $\sim 12(1000{\kms}/v_w)$ yr, taking a timescale of 120 days to the first feature.
\cite{kotak06} took a timescale of 150 days and derived a period for the mass loss
events of $\sim 25(200{\kms}/v_w)$ yr for SN 2003bg and SN 2001ig.
The main difference between these estimates is that \cite{kotak06} used an optically
determined maximum ejecta velocity of $15,000\kms$ at 14 days to determine the expansion of
the radio emitting region, whereas \cite{soderberg06} used the SSA model for the radio
emission to determine a mean velocity of about $40,000\kms$.
As discussed above and shown by the VLBI observations of SN 2008ax, the highest velocity
ejecta may not be detectable at optical wavelengths.
The variability timescales for SN 2008ax and SN 2008bo are not well defined, but appear
to be shorter than in the first two cases.

As noted by \cite{kotak06}, the timescale of decades is consistent with the variability
timescale of luminous blue variables (LBV) undergoing S Doradus-type variations and
they propose such stars as progenitors.
An example of such a star is AG Car, which has been followed from its minimum
(hotter) state through a cooler phase \citep{stahl01}.
In its minimum state, the mass loss rate is $\sim 3\times 10^{-5}\ml$ and
$v_w\approx 300\kms$, so that $A_*\approx 10$.
This is at the high end, but within the range, of wind densities discussed here
for the Type cIIb events.
However, the radius of AG Car in its compact state is $50\Rsun=3.5\times 10^{12}$ cm \citep{stahl01}.
This radius is a factor $\ga 10$ times
the radius deduced here for the progenitor of SN 2008ax and the radii of WR stars.
Also, \cite{stahl01} estimate a mass loss rate of $\sim 1.5\times 10^{-4}\ml$
and $v_w\approx 150\kms$ in the cool state, so that the wind density is $\sim 10$ times
higher than in the hot state.
The density contrast is higher if the faster wind sweeps the slower wind into a shell.
The shell densities are higher than appear present in the Type cIIb events and are
closer to the shell interaction observed in SN 2006jc \citep{foley07}.
S Doradus stars with lower luminosities than AG Car, which are more likely progenitors of the
objects discussed here, have even larger radii.

\cite{kotak06} note that a WR progenitor is possible if the star becomes a such a
star just before the supernova so it still interacted with mass loss from the LBV phase.
However this would require the transition $<25$ yr before the explosion.
In addition, the fast WR wind would modify the close in circumstellar medium.
Finally, the ejecta masses for SNe cIIb are not particularly large: estimates are
$2.5-5\Msun$ for SN 1996cb \citep{deng01},
$\sim 4\Msun$ for SN 2003bg \citep{mazzali09}, and
$\sim 3.8\Msun$ for SN 2008ax \citep{tsvetkov09}.
Unless a considerable amount of matter goes into a black hole central remnant,
these masses are less than expected for the cores of LBV stars.
Overall, the evidence points to progenitors for Type cIIb supernovae that
are less massive and more compact than LBVs, and are more
consistent with Wolf-Rayet stars.
Only a weak variation in wind properties is necessary to produce the observed radio
variations.
However, the possibility that the mechanism is similar to that producing
S Doradus variations in LBVs cannot by ruled out.

\section{DISCUSSION}

One expectation of the two types of IIb is that the Type cIIb events should have a lower
H mass envelope so that a large radial extent cannot be sustained.
For SN 1993J, \cite{woosley94} estimate a H mass of $0.2\pm 0.05\Msun$.
For SN 2003bg, \cite{mazzali09} estimate $0.05\Msun$ of H, in keeping with
expectations.
However, \cite{qiu99} suggest that SN 1996cb has more H than SN 1993J,
although \cite{deng01} estimate $0.1- 0.2\Msun$ of H if the explosion
energy is $10^{51}$ ergs.
The implied dividing line between the types is $\sim 0.1\Msun$.

For lower amounts of H, we expect that the Type cIIb objects merge into the
Type Ib's.
A recent example of a Type Ib with H is SN 2007Y which showed high velocity
($\sim 10,000\kms$) H absorption near maximum light \citep{stritz09}.
The finding of H in the spectra of Type Ib events is common \citep{branch02}.
\cite{branch02} estimate that a H mass $\sim 0.01\Msun$ is typically needed to produce
the required optical depth in H$\alpha$.
There may  not be a clear separation of the Type cIIb and Type Ib events, but
a gradual transition depending on remaining H mass.

The Cas A supernova was recently shown to be of Type IIb by the spectrum of its
light echo \citep{krause08}, which raises the question of whether the progenitor
was compact or extended.
Studies of the supernova remnant have not been conclusive; \cite{chevalier03}
argued that the progenitor was extended with the dense wind extending to the
stellar surface, but there have been arguments for a blue phase before the
supernova \citep{hwang09}.
The Type IIb identification was made on the basis of the close correspondence of the
echo spectrum with the time integrated spectrum of SN 1993J.  The H$\alpha$
line in the echo spectrum appears to be closer to the maximum light H$\alpha$
in SN 1993J than in SN 2003bg \citep{hamuy09}, but a more detailed comparison with
the time integrated spectrum of a Type cIIb event would be useful.

The existing observations that allow a distinction between Type cIIb and eIIb supernovae,
primarily at radio wavelengths, indicate a clear separation between the 2 types of
events, but the numbers are small.
Future time domain surveys should increase the discovery rate of Type IIb events so
that this issue can be resolved.
In addition, the early discovery of supernovae ($\la 2$ days from explosion)
will allow the classification of the events from their shock breakout
and cooling envelope emission.

\acknowledgments
We are grateful to the referee for a helpful report,
and acknowledge support from NSF grant AST-0807727 (RAC), a Hubble Fellowship (AMS),
and Swift GI program NNX09AR05G.

\clearpage

\begin{figure}[!hbtp]   
\epsscale{.80}
\plotone{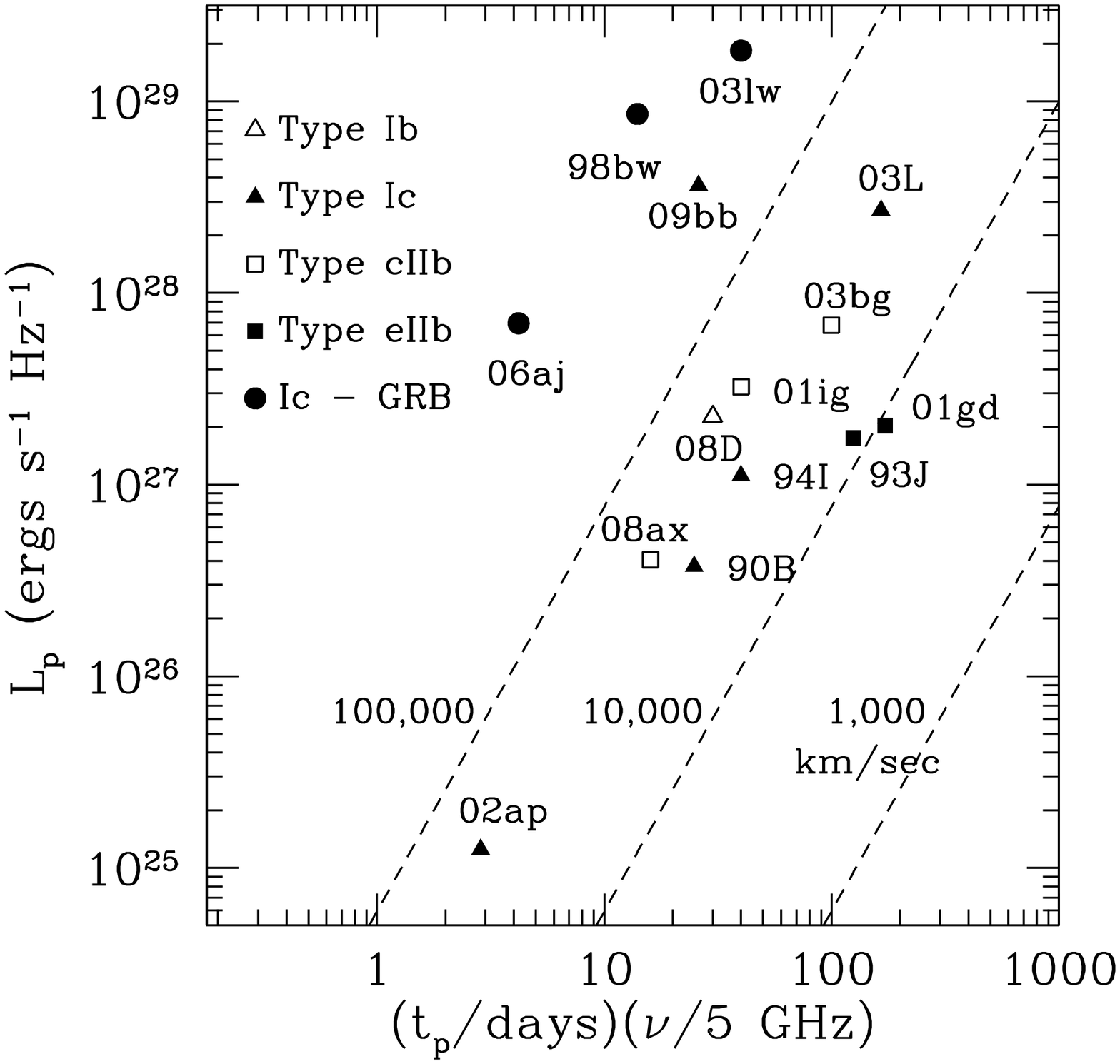}
\caption{Peak spectral radio luminosity vs. the product of the time of the
peak and the frequency of the measurement for supernovae of Types Ib/c and IIb.  
The observed supernovae are designated by the last two digits of the year and
letter(s).
References to the recent radio observations are \cite{soderberg08} 
for SN 2008D, \cite{roming09} for SN 2008ax, and \cite{soderberg09} for SN 2009bb;
other references can be found in \cite{chevalier06}.
The dashed lines show the mean velocity of the radio shell if synchrotron
self-absorption is responsible for the flux peak; a particle spectral index
$p=3$ is assumed \citep[see][]{chevalier98}.
}
\end{figure}

\end{document}